\documentclass[12pt,preprint]{aastex}

\newcommand{\bdv}[1]{\mbox{\boldmath$#1$}}

\def\au{{\rm AU}} 
 
\def\kms{{\rm km}\,{\rm s}^{-1}}
\def\masyr{{\rm mas}\,{\rm yr}^{-1}}
\def\kpc{{\rm kpc}}
\def\mas{{\rm mas}}
\def\muas{{\mu\rm as}}

\def\max{{\rm max}}
\def\rel{{\rm rel}}

\def\eff{{\rm eff}}
\def\hel{{\rm hel}}
\def\geo{{\rm geo}}
\def\e{{\rm E}}
\def\bpi{{\bdv\pi}}
\def\bmu{{\bdv\mu}}

\def\bv{{\bf v}}

\begin{document}
\title{{\it Spitzer} Parallax of OGLE-2015-BLG-0966: 
A Cold Neptune in the Galactic Disk}

\author{
R.A.~Street$^{R1}$,
A.~Udalski$^{O1}$,
S.~Calchi Novati$^{S1,S2,S3,a}$,
M.P.G.~Hundertmark$^{R2}$,
W.~Zhu$^{S4}$,
A.~Gould$^{S4}$,
J.~Yee$^{S5,b}$,
Y.~Tsapras$^{R3}$,
D.P.~Bennett$^{M1}$
}

\and

\author{The RoboNet Project \& MiNDSTEp Consortium,\\
U. G. J{\o}rgensen$^{R2}$,
M. Dominik$^{R4,c}$,
M.I. Andersen$^{R6}$,
E.~Bachelet$^{R1,R7}$,
V. Bozza$^{S2, R8}$,
D. M. Bramich$^{R7}$,
M. J. Burgdorf$^{R9}$,
A.~Cassan$^{R10}$,
S. Ciceri$^{R11}$,
G. D'Ago$^{S3}$,
Subo~Dong$^{R12}$,
D. F. Evans$^{R13}$
Sheng-hong Gu$^{R14}$, 
H.~Harkonnen$^{R2}$,
T. C.~Hinse$^{R15}$,
Keith~Horne$^{R4}$,
R. Figuera Jaimes$^{R4, R16}$,
N. Kains$^{R17}$,
E. Kerins$^{R17}$,
H. Korhonen$^{R19,R6}$,
M. Kuffmeier$^{R2}$,
L. Mancini$^{R11}$,
J.~Menzies$^{R20}$,
S.~Mao$^{R21}$,
N. Peixinho$^{R22}$,
A. Popovas$^{R2}$,
M. Rabus$^{R23,R11}$,
S. Rahvar$^{R24}$,
C.~Ranc$^{R10}$,
R. Tronsgaard Rasmussen$^{R25}$,
G. Scarpetta$^{S3, S2}$,
R. Schmidt$^{R3}$,
J. Skottfelt$^{R26}$,
C. Snodgrass$^{R27}$,
J. Southworth$^{R13}$,
I.A.~Steele$^{R28}$,
J. Surdej$^{R29}$, 
E. Unda-Sanzana$^{R22}$,
P. Verma$^{S3}$,
C. von Essen$^{R25}$,
J. Wambsganss$^{R3}$,
Yi-Bo. Wang$^{R14}$,
O. Wertz$^{R29}$
}

\and

\author{The OGLE Project,\\
R. Poleski$^{S4,O1}$,
M. Pawlak$^{O1}$,
M.K. Szyma{\'n}ski$^{O1}$,
J. Skowron$^{O1}$,
P. Mr{\'o}z$^{O1}$,
S. Koz{\l}owski$^{O1}$,
{\L}. Wyrzykowski$^{O1}$, 
P. Pietrukowicz$^{O1}$,
G. Pietrzy{\'n}ski$^{O1}$,
I. Soszy{\'n}ski$^{O1}$,
K. Ulaczyk$^{O2}$,
}

\and

\author{The Spitzer Team,\\
C. Beichman$^{S1}$,
G. Bryden$^{S6}$,
S. Carey$^{S7}$,
B. S. Gaudi$^{S4}$,
C. Henderson$^{S4}$,
R. W. Pogge$^{S4}$,
Y. Shvartzvald$^{S6,d}$
}

\and

\author{The MOA Collaboration,\\
F.~Abe$^{M2}$,
Y.~ Asakura$^{M2}$, 
A.~Bhattacharya$^{M1}$, 
I.A.~Bond$^{M3}$,
M.~Donachie$^{M4}$,
M.~Freeman$^{M4}$,
A.~Fukui$^{M5}$, 
Y.~Hirao$^{M6}$,
K.~Inayama$^{M7}$,
Y.~Itow$^{M2}$,
N.~Koshimoto$^{M6}$,
M.C.A.~Li$^{M4}$,
C.H.~Ling$^{M3}$,
K.~Masuda$^{M2}$,
Y.~Matsubara$^{M2}$,
Y.~Muraki$^{M2}$, 
M.~Nagakane$^{M6}$, 
T.~Nishioka$^{M2}$, 
K.~Ohnishi$^{M8}$,
H.~ Oyokawa$^{M2}$,
N.~Rattenbury$^{M4}$,
To.~Saito$^{M9}$,
A.~Sharan$^{M4}$,
D.J.~Sullivan$^{M10}$,
T.~Sumi$^{M6}$,
D.~Suzuki$^{M2}$,
P.,J.~Tristram$^{M11}$,
Y.~Wakiyama$^{M1}$, 
A.~Yonehara$^{M7}$
}

\and

\author{KMTNet Modeling Team,\\
C. Han$^{K1}$, 
J.-Y.Choi$^{K1}$,
H. Park$^{K1}$,
Y. K. Jung$^{K1}$, 
I.-G. Shin,$^{K1}$
}

\affil{$^{R1}$LCOGT, 6740 Cortona Drive, Suite 102, Goleta, CA~93117, USA}
\affil{$^{O1}$Warsaw University Observatory, Al.~Ujazdowskie~4, 00-478~Warszawa,Poland}
\affil{$^{S1}$NASA Exoplanet Science Institute, MS 100-22, California Institute of Technology, Pasadena, CA 91125, USA}
\affil{$^{S2}$Dipartimento di Fisica "E.R. Caianiello", Universit{\`a} di Salerno, Via Giovanni Paolo II 132, 84084, Fisciano, Italy}
\affil{$^{S3}$Instituto Internazionale per gli Alti Studi Scientifici (IIASS), Via G. Pellegrino 19, 84019 Vietri sul Mare (SA), Italy}
\affil{$^{R2}$Niels Bohr Institute \& Centre for Star and Planet Formation, University of Copenhagen, {\O}ster Voldgade 5, 1350 - Copenhagen K, Denmark}
\affil{$^{S4}$Department of Astronomy, Ohio State University, 140 W. 18th Ave., Columbus, OH 43210, USA}
\affil{$^{S5}$Harvard-Smithsonian Center for Astrophysics, 60 Garden St., Cambridge, MA 02138, USA ; Sagan Fellow.}
\affil{$^{R3}$Astronomisches Rechen-Institut, Zentrum f{\"u}r Astronomie der Universit{\"a}t Heidelberg (ZAH), 69120 Heidelberg, Germany}
\affil{$^{M1}$Department of Physics, University of Notre Dame, Notre Dame, IN 46556, USA}


\affil{$^{R4}$SUPA, School of Physics \& Astronomy, University of St Andrews, North Haugh, St Andrews KY16 9SS, UK}
\affil{$^{R5}$Astronomisches Rechen-Institut, Zentrum f{\"u}r Astronomie der Universit{\"a}t Heidelberg (ZAH), 69120 Heidelberg, Germany}
\affil{$^{R6}$Niels Bohr Institute \& Dark Cosmology Centre, University of Copenhagen, Juliane Mariesvej 30, 2100 - Copenhagen {\O}, Denmark}
\affil{$^{R7}$Qatar Environment and Energy Research Institute, Qatar Foundation, P.O. Box 5825, Doha, Qatar}
\affil{$^{R8}$Instituto Nazionale di Fisica Nucleare, Sezione di Napoli, Napoli, Italy}
\affil{$^{R9}$Meteorologisches Institut, Universit{\"a}t Hamburg, Bundesstra\ss{}e 55, 20146 Hamburg, Germany}
\affil{$^{R10}$Sorbonne Universit\'es, UPMC Univ Paris 6 et CNRS, UMR 7095, Institut d'Astrophysique de Paris, 98 bis bd Arago, 75014 Paris, France}
\affil{$^{R11}$Max Planck Institute for Astronomy, K{\"o}nigstuhl 17, 69117 Heidelberg, Germany}
\affil{$^{R12}$Kavli Institute for Astronomy and Astrophysics, Peking University, Yi He Yuan Road 5, Hai Dian District, Beijing 100871, China}
\affil{$^{R13}$Astrophysics Group, Keele University, Staffordshire, ST5 5BG, UK}
\affil{$^{R14}$Yunnan Observatories, Chinese Academy of Sciences, Kunming 650011, China}
\affil{$^{R15}$Korea Astronomy \& Space Science Institute, 776 Daedukdae-ro, Yuseong-gu, 305-348 Daejeon, Republic of Korea}
\affil{$^{R16}$European Southern Observatory, Karl-Schwarzschild Stra\ss{}e 2, 85748 Garching bei M\"{u}nchen, Germany}
\affil{$^{R17}$Space Telescope Science Institute, 3700 San Martin Drive, Baltimore, MD 21218, United States of America}
\affil{$^{R18}$Jodrell Bank Centre for Astrophysics, School of Physics and Astronomy, University of Manchester, Oxford Road, Manchester M13 9PL, UK}
\affil{$^{R19}$Finnish Centre for Astronomy with ESO (FINCA), V{\"a}is{\"a}l{\"a}ntie 20, FI-21500 Piikki{\"o}, Finland}
\affil{$^{R20}$South African Astronomical Observatory, PO Box 9, Observatory 7935, South Africa}
\affil{$^{R21}$National Astronomical Observatories, Chinese Academy of Sciences, 100012 Beijing, China}
\affil{$^{R22}$Unidad de Astronom{\'{\i}}a, Fac. de Ciencias B{\'a}sicas, Universidad de Antofagasta, Avda. U. de Antofagasta 02800, Antofagasta, Chile}
\affil{$^{R23}$Instituto de Astrof\'isica, Facultad de F\'isica, Pontificia Universidad Cat\'olica de Chile, Av. Vicu\~na Mackenna 4860, 7820436 Macul, Santiago, Chile}
\affil{$^{R24}$Department of Physics, Sharif University of Technology, PO Box 11155-9161 Tehran, Iran}
\affil{$^{R25}$Stellar Astrophysics Centre, Department of Physics and Astronomy, Aarhus University, Ny Munkegade 120, DK-8000 Aarhus C, Denmark}
\affil{$^{R26}$Centre for Electronic Imaging, Department of Physical Sciences, The Open University, Milton Keynes, MK7 6AA, UK}
\affil{$^{R27}$Planetary and Space Sciences, Department of Physical Sciences, The Open University, Milton Keynes, MK7 6AA, UK}
\affil{$^{R28}$Astrophysics Research Institute, Liverpool John Moores University, Liverpool CH41 1LD, UK}
\affil{$^{R29}$Institut d'Astrophysique et de G\'eophysique, All\'ee du 6 Ao\^ut 17, Sart Tilman, B\^at. B5c, 4000 Li\`ege, Belgium}


\affil{$^{O2}$Department of Physics, University of Warwick, Gibbet Hill Road, Coventry, CV4 7AL, UK}

\affil{$^{S6}$Jet Propulsion Laboratory, California Institute of Technology, 4800 Oak Grove Drive, Pasadena, CA 91109, USA}
\affil{$^{S7}${\it Spitzer}, Science Center, MS 220-6, California Institute of Technology,Pasadena, CA, USA}


\affil{$^{M2}$Solar-Terrestrial Environment Laboratory, Nagoya University, Nagoya 464-8601, Japan}
\affil{$^{M3}$Institute of Information and Mathematical Sciences, Massey University, Private Bag 102-904, North Shore Mail Centre, Auckland, New Zealand}
\affil{$^{M4}$Department of Physics, University of Auckland, Private Bag 92019, Auckland, New Zealand}
\affil{$^{M5}$Okayama Astrophysical Observatory, National Astronomical Observatory of Japan, 3037-5 Honjo, Kamogata, Asakuchi, Okayama 719-0232, Japan}
\affil{$^{M6}$Department of Earth and Space Science, Graduate School of Science, Osaka University, Toyonaka, Osaka 560-0043, Japan}
\affil{$^{M7}$Department of Physics, Faculty of Science, Kyoto Sangyo University, 603-8555 Kyoto, Japan}
\affil{$^{M8}$Nagano National College of Technology, Nagano 381-8550, Japan}
\affil{$^{M9}$Tokyo Metropolitan College of Aeronautics, Tokyo 116-8523, Japan}
\affil{$^{M10}$School of Chemical and Physical Sciences, Victoria University, Wellington, New Zealand}
\affil{$^{M11}$Mt. John University Observatory, P.O. Box 56, Lake Tekapo 8770, New Zealand}


\affil{$^{K1}$Department of Physics, Chungbuk National University, Cheongju 361-763, Republic of Korea}

\affil{$^{a}$Sagan Visiting Fellow}
\affil{$^{b}$Sagan Fellow}
\affil{$^{c}$Royal Society University Research Fellow}
\affil{$^{d}$NASA Postdoctoral Program Fellow}

\begin{abstract}
We report the detection of a Cold Neptune $m_{\rm planet}=21\pm 2\,M_\oplus$
orbiting a $0.38\,M_\odot$ M dwarf lying 2.5--3.3 kpc                   
toward the Galactic center as part of a campaign combining ground-based
and {\it Spitzer} observations to measure the Galactic distribution
of planets.  This is the first time that the complex real-time
protocols described by \citet{yee15}, which aim to maximize planet
sensitivity while maintaining sample integrity,  
have been carried out in practice.  Multiple survey and follow-up teams
successfully combined their efforts within the framework of these
protocols to detect this planet.  This is the second planet in the
{\it Spitzer} Galactic distribution sample.  Both are in the near-to-mid
disk and clearly not in the Galactic bulge.

\end{abstract}


\section{{Introduction}
\label{sec:intro}}

The 2015 {\it Spitzer} microlensing campaign is the first project with
the specific goal of characterizing the Galactic distribution of planets.
Such characterization has three requirements: 
\begin{enumerate}
\item[]{1) A survey that is sensitive to planets in substantially different
Galactic environments.}
\item[]{2) Well-characterized selection.}
\item[]{3) Ability to determine, at least statistically, the Galactic 
environment of each potential host, whether a planet is detected or not.}
\end{enumerate}

At present, microlensing is the only possible method to attack this
problem because all other planet search techniques fail criterion (1).  By
contrast, microlensing is about equally sensitive to planet-hosting lenses 
at all distances along the line of sight from the Sun to the Galactic-bulge sources
that are monitored for microlensing events.  

Initially microlensing planet searches were conducted in a fairly
opportunistic way, dominated by targeted follow-up of ``interesting''
events.  Because ``interesting'' was not necessarily well-defined, it
was difficult to satisfy criterion (2).  
Nevertheless \citet{gould10} and \citet{cassan12} 
were able to construct subsamples with well-characterized selection.  In parallel, the RoboNet and MiNDSTEp
teams developed robotic algorithms \citep{Horne2009, Dominik2010, Hundertmark2015} to carry out the target selection in a repeatable and
well-characterized manner.   With the advent of second generation surveys, particularly OGLE-IV and
KMTNet, uniform selection is becoming routine (see, e.g. \citealt{suzuki15,shvartzvald2015}). 
In their high-cadence zones, these wide field surveys can obtain dense enough
coverage that additional follow-up observations are not necessary to
detect and characterize planets. Hence, events can be monitored uniformly
with a pre-defined observing sequence.  

The main problem has always been to determine the location of
the lenses that are being probed for planets, including both those
in which planets are detected and those for which they are not.
For the
great majority of microlensing events, the lens (host) star is not definitively
identified and for many it is not detected at all.  For these cases,
the microlens parallax 
\begin{equation}
\bpi_\e\equiv {\pi_\rel\over\theta_\e}\,{\bmu\over\mu};
\qquad \theta_\e^2\equiv \kappa M\pi_\rel;
\qquad \kappa\equiv {4 G\over c^2\,\au}\simeq 8.1{\mas\over M_\odot},
\label{eqn:bpie}
\end{equation}
is the key to determining the lens mass and distance.  Here,
$\pi_\rel=\au(D_L^{-1}-D_S^{-1})$ and $\bmu$ 
are respectively the lens-source relative parallax
and proper motion, $\theta_\e$ is the angular Einstein radius, and
$M$ is the lens mass.  As detailed by Figure~1 from \citet{gouldhorne},
the amplitude $\pi_\e=\pi_\rel/\theta_\e$ is set by the fact that
motion by the observer of 1 AU displaces the apparent angular separation
of the lens and source by an angle $\pi_\rel$, which in turn induces
microlensing effects according to how large this displacement is
compared to the Einstein radius.  
In principle, there are two ways to observe the displacement caused by
parallax. First, the observer can wait to be moved by Earth's orbital motion, 
creating a displacement relative to simple rectilinear motion between the
source and lens. Second, two observers at substantially different locations
can compare their observations.

Some microlens parallaxes have been measured from the ground using the
first effect, particularly for long timescale events \citep{poindexter05}
and including fortuitously a significant number of microlens
planetary events \citep{gould10}. However, the sample of events with
parallaxes measured from the ground is extremely heavily biased toward
nearby lenses because they have projected Einstein radii
$\tilde r_\e\equiv \au/\pi_\e\sim 2$--$5\,\au$, roughly 3 times smaller
than is typical of bulge lenses. Hence, this
sample is almost unusable for measuring the Galactic distribution of
planets

The alternative is space-based parallaxes.  At $\sim 1\,\au$ from Earth,
{\it Spitzer} is ideally located to be such a ``microlens parallax satellite''.
It does, however, face a number of challenges.  First, due to
Sun-angle viewing restrictions, it can observe Galactic bulge targets
(which are near the ecliptic) for only 38 continuous days out of 
the 8 months they are visible from Earth.  Second, microlensing events
must first be detected from the ground and identified as reasonably
planet sensitive before they are targeted by {\it Spitzer}, and
these uploads occur 3--10 days before the observations begin
(Figure~1 of \citealt{ob140124}).  Since microlensing events often evolve quite 
rapidly, these practical constraints 
intrinsically restrict the pool of targets.  Finally,
{\it Spitzer} observes at $\lambda=3.6\,\mu$m, roughly 4.5 times
the wavelength of ground-based microlensing searches.  This creates
technical problems, some of which are described below.  See \citet{yee15}
for a more complete discussion.

In 2014 the Director granted 100 hours for a pilot program
to determine whether {\it Spitzer} could function effectively 
as a parallax satellite.  Special protocols were developed to
rapidly upload targets \citep{ob140124}. New techniques 
were developed by \citet{ob140939} and \citet{21event}
to break the famous ``4-fold degeneracy'' \citep{refsdal66,gould94}
that had been believed to require extremely aggressive observational
strategies (e.g., \citealt{gould95,gaudi97}). For the two events
with measured (OGLE-2014-BLG-1050, \citealt{ob141050}) or strongly constrained
(OGLE-2014-BLG-0124, \citealt{ob140124})
$\theta_\e$ (using the standard technique for
events with caustic features, \citealt{ob03262}), the lens mass $M$
and distance $D_L$ were determined using inversions of 
Equation~(\ref{eqn:bpie}),
\begin{equation}
M={\theta_\e\over \kappa\pi_\e};
\qquad \pi_\rel = \theta_\e\pi_\e;
\qquad D_L ={\au\over \pi_\rel + \pi_S}
\label{eqn:massdis}
\end{equation}
Here $\pi_S$ is the source parallax which is
almost always well understood because the source is visible. More critically,
since the overwhelming majority of single-lens events do not show
caustic features, a robust method was developed for 
estimating distances kinematically for events with measured $\bpi_\e$
but not $\theta_\e$ \citep{21event}.

However, the 2014 {\it Spitzer} campaign was entirely focused on
demonstrating the feasibility of the method, and no systematic
effort was made to find planets. Nevertheless, one planet was discovered
\citep{ob140124}.

The successes of the 2014 campaign and the breakthroughs it precipitated
led to the realization that satellite parallax measurements could be used
to measure the Galactic distribution of planets. In 2015, 832 hours were
awarded for a program whose primary objective was to do just that
\citep{spitzer2015prop}.
In fact, as
specifically argued in the proposal, several such annual campaigns will
be required to acquire sufficient statistics to make this measurement.

As discussed in detail by \citet{yee15}, the observational protocols
required to
\begin{enumerate}
\item[]{1) maximize the sensitivity of the survey to planets, while}
\item[]{2) maintaining well-characterized selection}
\end{enumerate}
are both intricate and complex.  We describe in some detail how
these applied to the case of the planetary detection reported below.
However, those readers interested in a full understanding must
actually study \citet{yee15}.

Here, we report the discovery and characterization of the planet
OGLE-2015-BLG-0966Lb. This is the second planet 
(after OGLE-2014-BLG-0124Lb, \citealt{ob140124})
in the statistical sample of microlens-parallax planets that can be
used to determine the Galactic distribution.  The observations, and
the observational protocols that guided them are described in 
Section~\ref{sec:obs}.  The ground-based and {\it Spitzer}
lightcurves are combined to yield microlensing parameters for the event
(including the microlens parallax $\bpi_\e$) in Section~\ref{sec:anal}.
There are some challenges in the estimate of $\theta_\e$ relative
to the usual case, which are discussed in Section~\ref{sec:thetae}.
After resolving these, we present the system physical parameters in
Section~\ref{sec:physical}.  Finally, we discuss some implications
of this discovery in Section~\ref{sec:discuss}.


{\section{Observations}
\label{sec:obs}}

\subsection{OGLE Alert and Observations}

All 2015 {\it Spitzer} targets were chosen based on microlensing
alerts issued by the Optical Gravitational Lens Experiment (OGLE)
or the Microlensing Observations in Astrophysics (MOA) collaboration,
with a substantial majority coming from OGLE.

The OGLE alert for OGLE-2015-BLG-0966 was issued on 2015 May 11, well
in advance of the {\it Spitzer} campaign, for which the first observations
were on June 6.   It lies at equatorial coordinates 
(17:55:01.02, $-29$:02:49.6), with Galactic coordinates $(0.96,-1.82)$, 
placing it in OGLE field BLG505, which implies that it is observed
at 20 minute cadence with OGLE's 1.3m Warsaw Telescope at the 
Las Campanas Observatory in Chile \citep{ogleiv}.
These dense observations permitted
OGLE to alert the event, using its Early Warning System (EWS) real-time 
event detection software \citep{ews1,ews2}, when the source had just
entered the Einstein ring, and so was just 0.38 mag above its $I=19.62$
baseline. OGLE continued to observe with this cadence throughout the
event except for interruptions due to weather and monthly passage of the
Moon through the bulge.

\subsection{{\it Spitzer} Observations}

A key element in meeting Criterion (2) for
measuring the Galactic distribution of planets is that events must be
selected for {\it Spitzer} observations without allowing any knowledge about
the presence or absence of planets to influence that decision.
As described in \citet{yee15}, {\it Spitzer} targets can be chosen
``objectively'' or ``subjectively''.  If they meet certain specified
criteria as of 6 hours before target submission (Mondays at UT 15:00),
then they are objective and {\it must} be chosen for observations with a certain
specified cadence.  In this case, all the planets discovered (as well
as all planet sensitivity, i.e., ability to detect planets) 
whether from before or after the
{\it Spitzer} observations begin, can be incorporated into the analysis.
In addition, if the event has not yet been selected objectively, the team may
{\it at any time}
still choose the event ``subjectively'' guided qualitatively by the goal of
maximizing the sum (over all choices) of the products $\sum_i S_iP_i$
where $S_i$ is the sensitivity of event $i$ to planets and $P_i$ is the probability
that a microlens parallax will actually be measured.  Only data taken
(or rather, made public)
after this selection date may be considered when calculating $S_i$.
The cadences of
subjectively chosen events can also be specified subjectively, but as
a practical matter they are usually specified to be the same as for
objectively chosen events.  Note that exclusion of 
{\it Spitzer} data from the analysis applies only to determining
whether or not an event enters the sample.  Once the event has a sufficiently
well measured parallax to enter the sample, 
all {\it Spitzer} data can be applied to improve the precision of the parallax
measurement or to search for planets.

If there remains time for additional {\it Spitzer} observations 
after scheduling all targets according to their adopted cadences,
this is applied to high-magnification events, with cadence ranked
according to predicted magnification in the observation interval.

A point of direct relevance to the present case is that if an event
is initially chosen subjectively but later meets objective criteria,
the objective status takes precedence.  This means that all planets
and planet-sensitivity from before the subjective alert can now be
included in the analysis, but also that only observations following
the objective alert can be used to determine if the parallax is measured
well enough to enter the sample.  If these objectively-based parallax data
are not adequate for a measurement, then all the {\it Spitzer} data can
be included, but then only planets and planet sensitivity from after
the subjective alert can enter the analysis.

Figures 2 and 3 from \citet{yee15} show flowcharts for this decision-making
process.

On Monday June 15, the {\it Spitzer} team chose OGLE-2014-BLG-0966 for
``secret'' observations at 1-day cadence.  
The purpose of ``secret'' observations is to resolve the tension between two
aspects of the observations. First, any event selected for {\it Spitzer}
observations must be observed and continue to be observed at some
predetermined cadence for a predetermined amount of time. However, {\it Spitzer}
targets can only be updated once per week. Hence, if the future course of
an event is uncertain at the time the targets must be sent to {\it Spitzer}
for observations, it may be selected as ``secret.'' Until it is formally
selected, none of the {\it Spitzer} observations may be included in the
calculation of parallax but if the event turns out to have low planet
sensitivity and/or poor chances of a viable parallax measurement, no
commitment has been made to continue observing it in the future, which
would be a waste of resources.
Hence, ``secret'' observations permit the team to defer this choice
until more information about the event is available, with a maximum
``loss'' of one week's observations.  In this case, however, the
team gained sufficient confidence in the event just two days later
to announce it publicly on 2015 June 15 UT 21:31.  The cadence was
specified as ``objective'', meaning once per day, which could not
then be altered.

However, the next week, on Monday June 22, the event qualified for
bonus observations because it was predicted to be high magnification
(as seen from Earth) during that week of {\it Spitzer} observations,
June 24 UT 11:43 to July 1 UT 16:46.  According to the prescriptions 
of \citet{yee15} all observing time that is not allocated for regular
observations should be applied to high-magnification events, rank
ordered by the $1\,\sigma$ lower limits of their predicted peak
magnification during the observing interval.  This prediction is
for magnification at Earth if that is all the information available
(as was the case for this event in that week).  OGLE-2015-BLG-0966
was estimated as $A_\max>11$ and so was assigned 8-per-day cadence.
It received a total of 58 observations during that week.  In fact, this
week did prove to contain the event's $A_\max\sim 100$ peak as seen
from the ground.  The
magnification of the event during the next week was less than the
cutoff for distributing extra Spitzer observations, so it was observed at
the previously determined cadence of 1/day.
This week it also met the objective
criteria for selection of rising events (criteria ``B'' in \citealt{yee15}).
This meant that if the event parallax could be adequately measured using
{\it Spitzer}
data from after July 1 UT 16:46, then all planets and planet sensitivity
could be included. Otherwise only planets and sensitivity from after
2015 June 15 UT 21:31 could be included (assuming all {\it Spitzer}
data were enough to measure a microlens parallax).  For the final ``week''
(actually about 10 days), the event again met the criteria for
additional high-magnification (as seen from Earth) observations and
was slated for a cadence of 4-per-day.  It may seem strange that
it would fail these criteria on the fifth week but pass them on the
sixth week, since the event was falling during this time.  However,
the criteria for receiving extra observations due to high-magnification
are (necessarily) based on predicted magnification,
not actual magnification, and these predictions improve with time.
Second, the competition from other events varies each week, so there
is strict correspondence between (predicted) magnification and observations
only within a given week, not between weeks.

Altogether, {\it Spitzer} observed this event a total of 129 times, each
with 6 dithered 30 s exposures.

\subsection{MOA data}

MOA independently identified this event on 16 June and monitored
it as MOA-2015-BLG-281 using its 1.8m telescope with $2.2\,\rm deg^2$
field at Mt.\ John New Zealand.  In contrast to most other
observatories, which observe in $I$ band, MOA observes in a broad
$R$-$I$ band pass.  The MOA cadence for this field is 15
minutes.  Thus, during the long June-July nights, OGLE and MOA
together nominally cover this field for about 18 hours per night.
However, the weather in New Zealand is far worse than in Chile, so
the fraction of nights that actually have this near-continuous coverage
is under 50\%.

\subsection{Ground-Based Follow-up Data}

\subsubsection{Follow-up Strategy}

As discussed in \citet{yee15}, ground-based follow-up strategy is intimately
connected with event selection, both objective and subjective.
Approximately 150 (nominally) point-lens events were selected for
{\it Spitzer} observations during the 2015 campaign, which far
exceeds the resources of all follow-up groups combined to densely
monitor events to search for planets.  This tension has two interrelated
implications.  First, follow-up groups are explicitly encouraged
{\it not} to monitor events that are heavily monitored by surveys,
with survey coverage being described in some detail at the time
of announcement of each event.  Second, there is a strong bias for the
{\it Spitzer} team to select events that are heavily monitored by surveys, 
exactly
because coverage is not dependent on limited follow-up resources.  Indeed, for
events (such as OGLE-2015-BLG-0966) that have 20-minute OGLE cadence,
there is an extremely strong bias because additional follow-up from the 
same time zone would be redundant, and hence the follow-up observing
resources are best applied to other events without such intense survey 
coverage.
Similarly for events in the $16\,\rm deg^2$ core
KMTNet fields, which have roughly 15-minute cadence from three observatories,
these considerations apply even more strongly.  OGLE-2015-BLG-0966
lies approximately in the middle of one of the four KMTNet prime fields.
However, the {\it Spitzer} team's map of these fields was precise
enough to recognize that the event lay in a gap between chips, so
that it was not actually covered by KMTNet.  (KMTNet did not at this
time publicly list the events that it was monitoring).

As OGLE-2015-BLG-0966 approached its peak (HJD$^\prime$ 7205.2, July 1.7), 
the very high cadence and
high quality of OGLE data permitted an accurate estimate of the
peak magnification $A_\max\sim 90$, which would make the event
highly sensitive to planets \citep{griest98}.  Because OGLE would
normally densely cover the event during the long Chile night,
while MOA would cover the even longer New Zealand night, the
need for follow-up would have appeared minimal.  This is particularly
true for the fairly large number of follow-up telescopes in Chile,
whose observations would be completely redundant with OGLE.

However, the peak of this event happened to occur when the Moon
was passing through the bulge, during which time OGLE does not observe.
Hence, several follow-up groups concentrated their efforts on this
event.  Nevertheless, the event did not gain the undivided attention
of follow-up groups due to several competing events.  Most notable
was OGLE-2015-BLG-0961, which peaked less than 12 hours earlier and at
even higher magnification $A_\max=200$.  However, this event was covered
by KMTNet and so assumed much lower (but still not zero) priority
from follow-up groups.

In particular, we note that the planet was discovered only because
follow-up groups recognized that there would
be no survey coverage of this event over Chile.

\subsubsection{LCOGT}

Las Cumbres Observatory Global Telescope Network (LCOGT) provided
ground-based observations primarily from its
southern ring of 8 1.0m telescopes sited at CTIO/Chile, SAAO/South
Africa and Siding Spring/Australia
\citep{Brown2013}. Two telescopes in Chile are equipped with the new
generation of Sinistro imagers that
incorporate 4k$\times$4k Fairchild CCD-486 Bl CCDs and offer a field of view
of $27^\prime\times 27^\prime$.  All other 1.0m
telescopes support SBIG STX-16803 cameras with Kodak KAF-16803 front
illuminated 4096x4096pix CCDs, used in bin
2$\times$2 mode with a field of view of $15.8^\prime\times 15.8^\prime$.
Observations were also made from the 2.0m
Faulkes North Telescope in Haleakala, Hawaii, using its
$10^\prime\times 10^\prime$ Spectral camera (Fairchild
CCD-486 Bl CCD).  All telescopes in the network are equipped with a
consistent set of filters.  SDSS-$i^\prime$
was used for the large majority of these observations, with a small
number made using the Bessell-$V$ filter.
Due to the constraints described above, LCOGT employed its TArget
Prioritization (TAP) algorithm
\citep{Hundertmark2015} to select a sub-set of events from the Spitzer
target list based on their predicted 
sensitivity to planets, which were drawn from {\it Spitzer} targets that 
fell in regions of
lower survey observing cadence.  Since
OGLE-2015-BLG-0966 falls in a region of high survey cadence, it was
not selected for observation until the
Moon's passage through the Bulge interrupted survey observations.  At
that point the event was flagged for high
density observations on both 1.0m and 2.0m networks.  These
observations were conducted as groups of 2--10
exposures repeated at intervals of a few minutes to hours, with less
dense observations being taken as the
event returned to baseline.  Hence, the data typically have dense
packets of coverage followed by short gaps.

\subsubsection{Danish Telescope}

The Danish 1.54m telescope is one of the national telescopes hosted
by ESO at La Silla in Chile. After a
successful refurbishment in 2012 by the Czech company Projectsoft, it
was equipped with the first routinely
operated multi-color instrument providing Lucky Imaging photometry
\citep{Skottfelt2015}.  The instrument
itself consists of two Andor iXon+ 897 EMCCDs and two dichroic mirrors
splitting the signal into red and 
visual bandpasses. For the 2015 microlensing campaign the camera was
operated at 10 Hz, and lucky
exposures were calibrated and tip-tilt corrected as described by
\citet{Harpsoe2012}.  Photometry was obtained from the collapsed images
using the DanDIA pipeline
\citep{Bramich2008} and based on routines of the
RoboNet reduction pipeline.   A modified version of ARTEMiS \citep{Dominik2008} was deployed to coordinate the observation of microlensing targets, and in the case of OGLE-2015-BLG-0966 resulted in confirmation of the anomaly from a second site.

\subsubsection{CTIO-SMARTS}

For the duration of the {\it Spitzer} campaign,
the Microlensing Follow Up Network ($\mu$FUN) doubled its normal
allocation to 6 hours per night on the dual optical/IR ANDICAM
camera on the 1.3m SMARTS telescope at CTIO.  This facility was
tasked with several objectives that were not always completely compatible.
These included regular monitoring of {\it Spitzer} targets in low-cadence
OGLE fields in order to predict their future behavior, sparse monitoring
of all targets in order to measure their $H$-band source fluxes
using ANDICAM's IR channel, and dense monitoring of events that
were at fairly high or very high magnification in order to detect
planets.  Hence, similar to the LCOGT telescopes, CTIO-SMARTS observed
in blocks with short gaps during which other goals (including other
science and shared resources) were pursued.

\subsection{Other Bands}

OGLE, LCOGT, and CTIO-SMARTS all obtained $V$-band observations in order
to measure the $V-I$ source color in order to characterize the
source.  As described above, CTIO-SMARTS automatically obtained
$H$-band observations through the IR channel simultaneously
with both the $V$-band and $I$-band optical observations.  These
also are used for source characterization and are not included in the
analysis.

\subsection{Data Reduction}

All ground-based data were reduced using standard algorithms.
All data entering the main analysis
used variants of image subtraction \citep{alard98, Bramich2013}.
CTIO-SMARTS used DoPhot \citep{dophot} reductions for the 
source-characterization analysis.

As explained in \citet{yee15}, no previously existing {\it Spitzer}
reduction software was suitable for variable stars in crowded fields.
Therefore customized software had to be developed.   This will
be described in a forthcoming work (Calchi Novati et al.\ 2015, in prep).

{\section{Lightcurve Analysis}
\label{sec:anal}}

Real-time lightcurve analysis, both manual and automated
\citep{Bozza2010} was conducted during the event, and immediately
alerted observers to the presence of the anomaly.  Extensive offline
analysis was subsequently carried out once data gathering on the event
was complete.

The analysis reported below is based on a simultaneous fit to
ground-based and {\it Spitzer} data.  However, the key event
characteristics can be most easily understood by considering
the two lightcurves separately, and indeed many characteristics
can be inferred by visual inspection.

\begin{figure}
\plotone{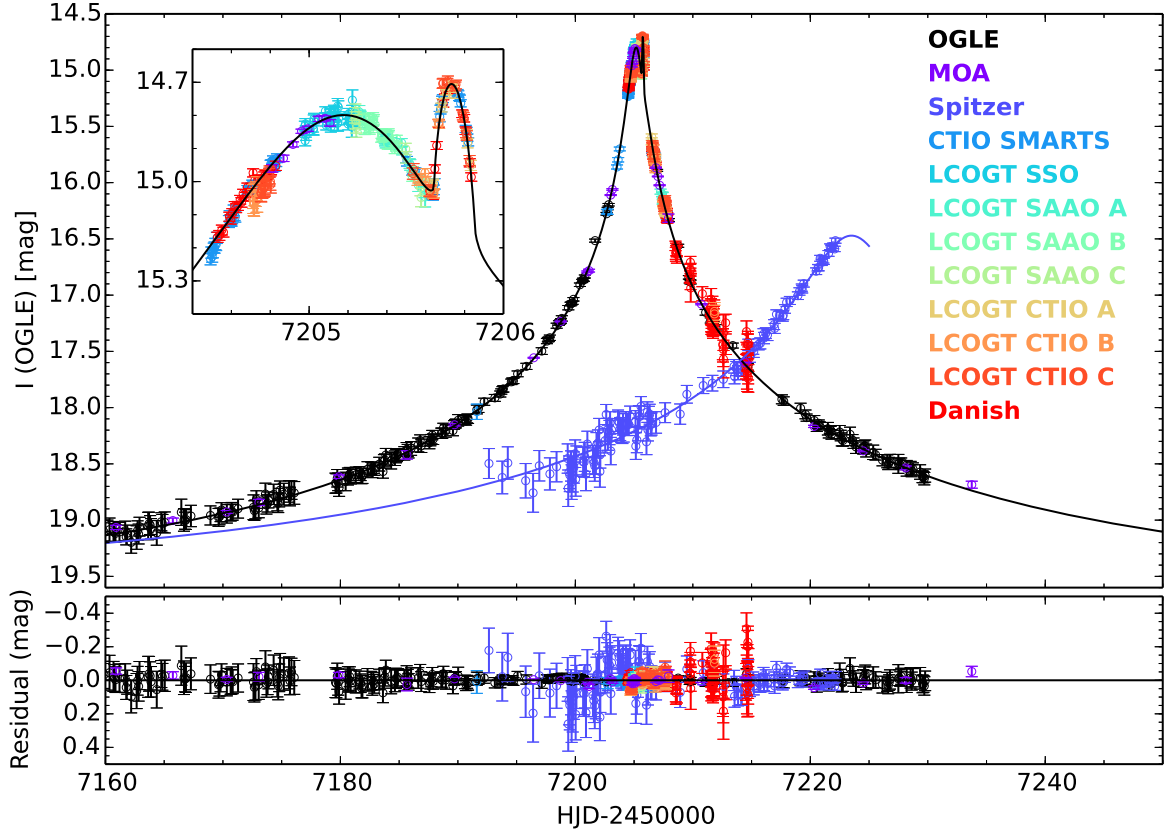}
\caption{OGLE-2015-BLG-0966 lightcurve.  Combined data
from 10 telescopes (color coded) trace the ground-based light curve
nearly continuously.  With the exception of the $\sim 6\,$hr post peak
bump, it is well characterized by a high-magnification point-lens
\citet{pac86} curve.  As described in the text, five of the seven
parameters $(t_0,u_0,t_\e,\alpha,\rho)$ needed to describe this curve
can be read off lightcurve or extracted with very simple analysis.
The remaining two $(s,q)$ (planet-star separation and mass ratio)
require more detailed modeling.  The {\it Spitzer} lightcurve is
aligned to the OGLE scale so that equal ``magnitudes'' represent
equal magnifications.  The microlens parallax $\pi_\e$ can be
well-estimated simply by comparing the {\it Spitzer} and ground-based
light curves.  See Section~\ref{sec:anal}.  (OGLE and MOA data are
binned for display in the Figure, but not in the fit.  Ground-based
data points with
uncertainties $>0.2$ mag are suppressed in the figure to avoid clutter,
but are included in the fit.)
}
\label{fig:lc}
\end{figure}

Ignoring for the moment the post-peak perturbation centered at
$t_{\rm pert}=7205.75$, the ground based lightcurve shown 
in Figure~\ref{fig:lc} peaks at $t_0=7205.19$ almost 4.85 mag above baseline,
indicating a high-magnification event $A_\max\ga 85$ (i.e., more if the
baseline source is blended).  In the neighborhood of the peak,
such events are characterized by flux evolution 
$F(t) = F_{\rm peak}(1 + (t-t_0)^2/t_\eff^2)^{-1/2}$ where $t_\eff=u_0 t_\e$,
$t_\e$ is the Einstein timescale and $u_0$ is the impact parameter
normalized to $\theta_\e$ \citep{gould96}.  Inspection of the peak region shows
$t_\eff = 0.68\,$days.  (For example, CTIO data at $\pm 1.50$ days from
peak are 0.96 mag below peak.  Hence, 
$t_\eff = 1.5\,{\rm day}(10^{0.8*0.96}-1)^{-1/2} = 0.68\,$day.)\ \
The fact that there is no pronounced dip just
prior to the abrupt rise due to a caustic crossing at $t_{\rm cc} = 7205.64$
shows that the companion/host mass ratio is very small, i.e., this
is a planetary rather than binary system.  Hence, the center of mass
of the system is very close to the ``center of magnification''.  It 
is the source motion relative to the ``center of magnification''
that produces the overall \citet{pac86} curve over the
unperturbed portions of the
light curve.  Hence, the caustic that induces the perturbation, and
which lies along the planet-host axis, points toward this center of
magnification.  This implies that the source is moving at an angle
$\alpha=\tan^{-1}(t_\eff/(t_{\rm pert}-t_0))=0.88\,$radians ($50.5^\circ$)
relative to the planet-host axis.  There is no dip between the caustic
entrance and exit, which implies that the caustic edges are separated
by significantly less than the source radius $\rho$ (normalized to $\theta_\e$).
Hence, the source radius crossing time, $t_*\equiv\rho t_\e$, can be estimated
$t_*\simeq (t_{\rm cc}-t_{\rm pert})*\sin(\alpha) = 0.07\,$days.  While
it is not obvious by inspection, a simple point-lens
fit to the unperturbed lightcurve shows that the source is essentially
unblended, so that in fact $A_\max = 85$, so $u_0=0.012$, and hence
$t_\e=t_\eff/u_0 = 57\,$days, and thus $\rho=1.2\times 10^{-3}$.  That is,
$(t_0,u_0,t_\e,\alpha,\rho)=(7205.19,0.012,57\,{\rm day},51^\circ,
1.2\times 10^{-3})$.
These by-eye estimates agree reasonably well with the parameter values derived from the best-fitting models (described below),  presented in 
Tables~\ref{tab:all} and \ref{tab:brief}.  
The remaining two parameters that can be extracted from the ground-based
lightcurve, i.e., the planet-star mass ratio $q$ and the planet-star
separation $s$ (normalized to $\theta_\e$), cannot be estimated by eye
and must be determined from the fit.

The effect of adding the {\it Spitzer} data is to determine the microlens
parallax $\bpi_\e$, which is incorporated directly into the fit in 
equatorial (north, east) coordinates.  However, as illustrated in
Figure~1 of \citet{gould94}, the impact of these measurements is
better visualized in the frame defined by the projected {\it Spitzer}-Earth
axis ${\bf D_\perp}$, in which
\begin{equation}
\bpi_\e = {\au\over D_\perp}(\Delta\tau,\Delta\beta);
\qquad \Delta\tau = {t_{0,\oplus} - t_{0,\rm sat}\over t_\e};
\qquad \Delta\beta = \pm u_{0,\oplus} - \pm u_{0,\rm sat},
\label{eqn:pieframe}
\end{equation}
and where the subscripts indicate parameters as measured from Earth
and {\it Spitzer}.

For the intuitive analysis of the {\it Spitzer} lightcurve (Fig.~\ref{fig:lc})
we begin with the external information
that the timescale $t_\e\simeq 57\,$days is known from the ground-based
lightcurve.  In fact, since Earth-{\it Spitzer} motion (projected on the sky)
is about $20\,\kms$ at the time that we will evaluate 
Equation~(\ref{eqn:pieframe}), the timescales will not be exactly the same.  
However,
this is expected to be a relatively small effect.  Hence, we
ignore it here with the proviso that we will later check for self-consistency.
The {\it Spitzer} lightcurve shows a factor $\sim 4$ rise over the 30
days of observations ending at 7222.14, where it is clearly turning over,
implying a peak $t_{0,\rm sat}\sim 7225$.  Thus, the interval between
peaks, normalized to the Einstein timescale is 
$\Delta\tau =(t_{0,\oplus}-t_{0,\rm sat})/t_\e \sim 0.35$.
Since 30 days before the peak, $u_{\rm sat}\sim 0.5$ ($A\sim 2$), 
the magnification at
peak is at least $A_{\max,\rm sat}\ga 8$ (more if the 
{\it Spitzer} source flux is blended),
so $|\Delta\beta| = |u_{0,\oplus}-u_{0,\rm sat}|\simeq u_{0,\rm sat}\la 0.12$.
{\it Spitzer} was 1.39 AU from Earth midway between the peaks, 
with $D_\perp=1.27\,\au$ after taking account of projection.  Since
it is approximately due West of Earth, the two coordinates in
Equation~(\ref{eqn:pieframe}) basically correspond to East and North,
respectively.  Hence, we derive $\pi_{\e,E}\sim -0.28$, and $|\pi_{\e,N}|<0.1$
As a final check, we note that these values imply a projected velocity
in the Earth frame of $\tilde v = \au/(\pi_\e t_\e)\sim 100\,\kms$,
meaning that Earth's motion is not completely negligible.  In principle,
we could make recursive corrections, but the main point of this exercise
is to show that most of the results of the full light curve analysis
can be basically derived by simple inspection of the {\it Spitzer}
and ground-based lightcurves.

We systematically model the combined Earth-based and {\it Spitzer}
lightcurves, notwithstanding the above analysis showing that five of the
seven Earth-based parameters are well-determined by visual inspection
and simple analysis. We conduct a systematic search of parameter space
using two different techniques, one based on an $(s,q,\alpha)$ grid and the
other based on lightcurve morphologies.  Models are evaluated based on the reduced $\chi^{2}$ of the fit per dataset ($\chi^{2}_{\rm{red}}$).  To account for variations in the estimation of photometric errors in different datasets, we adopt the common practise (e.g. \citet{bachelet2012}) of re-normalize the errors according to:

\begin{equation}
e_{\rm{new}} = a_{0}\sqrt{ e^{2}_{\rm{orig}} + a^{2}_{1} }
\end{equation}

Coefficients $a_{0}$ and $a_{1}$ are set such that the $\chi^{2}_{\rm{red}}$ of each dataset relative to the model equal unity.

We found that the only viable solutions do in fact have the five parameters as approximately specified above.
Moreover, $s$ and $q$ are well localized, except that (as is often the
case) there is a degenerate solution $s\rightarrow s^{-1}$ \citep{griest98}.

\begin{figure}
\plottwo{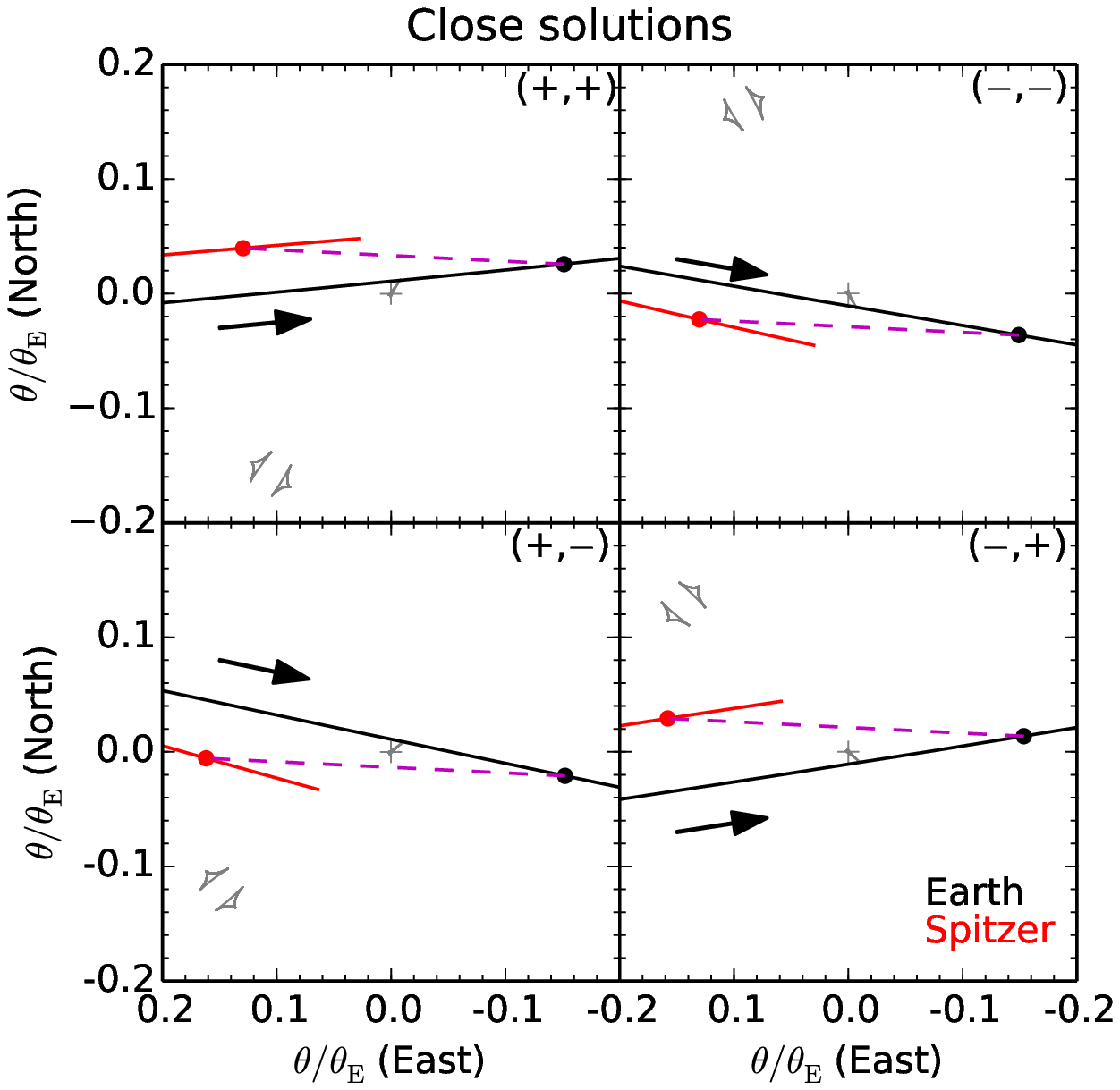}{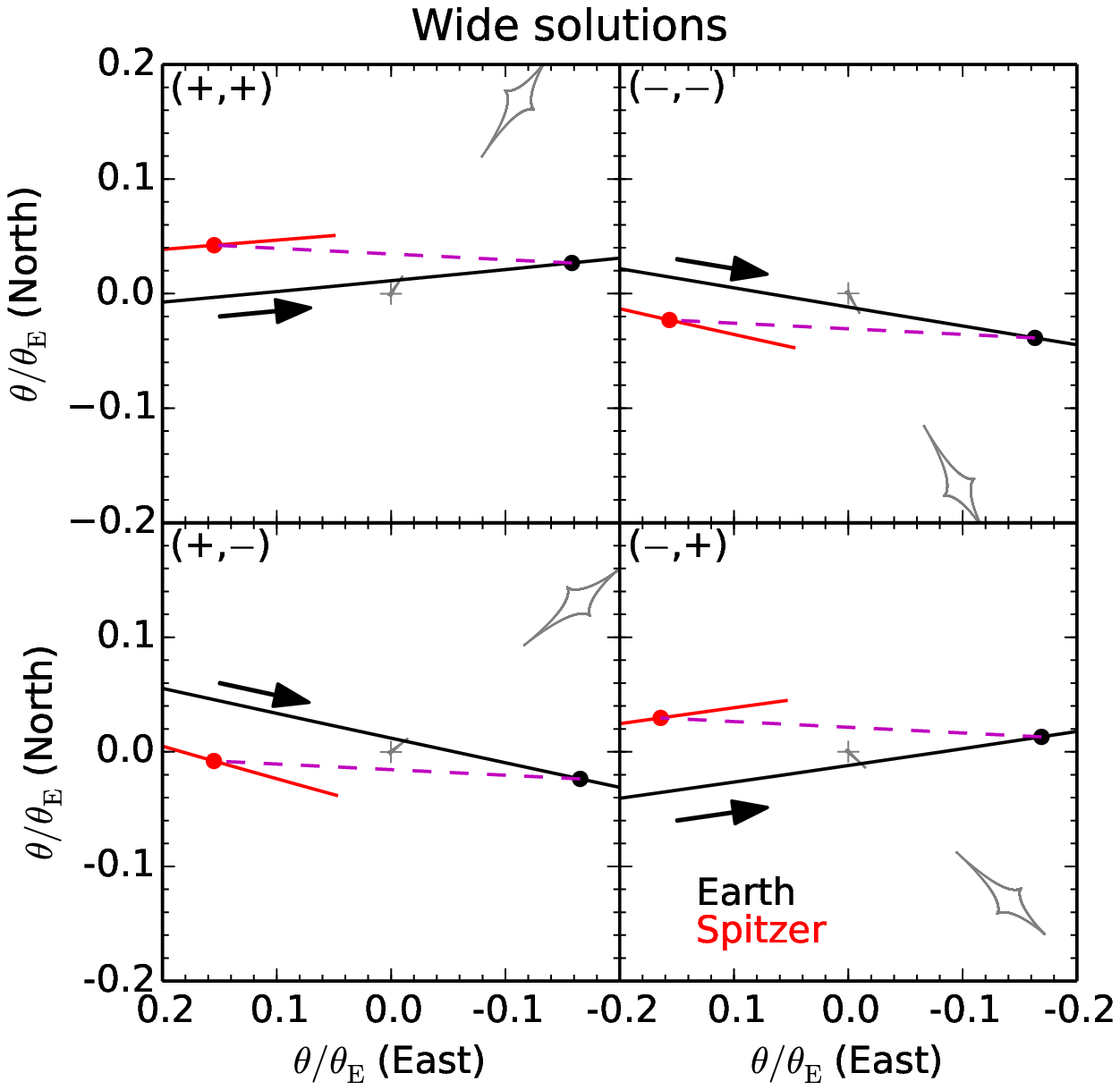}
\caption{Thumb-nail sketches of all eight caustic geometries
for planetary microlensing event OGLE-2015-BLG-0966.
The black and red lines represent the trajectories as seen from 
Earth and {\it Spitzer}, respectively.  The right ends of the red lines
represent the end of {\it Spitzer} observations at HJD 7222.14.
All eight solutions have essentially the same physical
implications.  First, while the overall caustic structures (closed black curves)
differ between close (left) and wide (right) binary solutions, the central
caustics, which are the only part probed by the Earth and {\it Spitzer}
observations, are virtually identical.  These are related by the 
$s\leftrightarrow s^{-1}$ degeneracy, and since $s\simeq 1.1$ is close to
unity, they correspond to physically similar systems.  The microlens
parallax $\bpi_\e$ is essentially determined by the offset between
{\it Spitzer} and Earth trajectories at the same epoch
(magenta dashed line segments).  See Equation~(\ref{eqn:pieframe}).
The amplitude of $\bpi_\e$ does
differ slightly between the $(\pm,\pm)$ and $(\pm,\mp)$ solutions because
the source is on the same side of the lens as seen from Earth and {\it Spitzer}
for the former and on the opposite side for the latter, leading to a different
distance perpendicular to the direction of motion.  However, because the
event is high magnification as seen from Earth ($u_{0,\oplus}\ll 1$), this
difference is itself small, and because the
separation along the direction of motion is much larger than the separation
perpendicular, this has almost no effect on the magnitude of the parallax,
$\pi_\e$, which is what enters the main physical parameters.  
See Table~\ref{tab:all}.
The only substantial difference among these eight solutions is that the motion
is somewhat north of west for the $(+,\pm)$ solutions and somewhat south
of west for the $(-,\pm)$ solutions.  However, this 
small difference has
no impact on the main physical parameters, which do not depend on this
direction.
}
\label{fig:caust1}
\end{figure}

\begin{figure}
\plotone{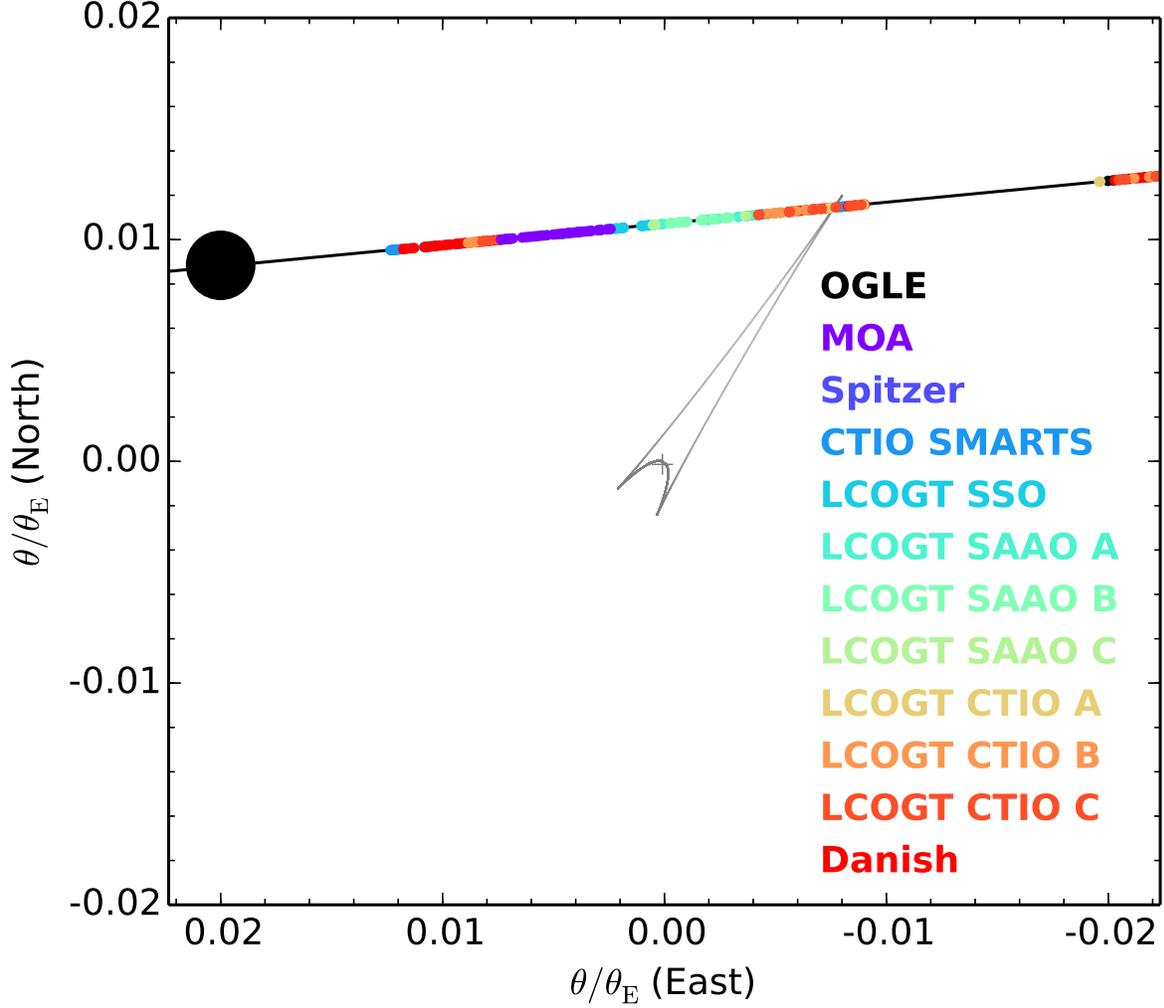}
\caption{Detail of caustic structure for one of the eight degenerate solutions
shown in Figure~\ref{fig:caust1} (close $(-,-)$).  
The colored circles represent the
data points on the light curve shown in Figure~\ref{fig:lc}.  The black filled circle on the trajectory represents the size of the source.  As described
in Section~\ref{sec:anal}, the angle between source trajectory and
the binary axis (same as the caustic axis) is determined directly by the
time offset of the anomalous peak from the overall peak, relative
to the curvature of the peak, without reference to any model.  However,
the derivation of $(s,q)$, which determines the overall caustic structure
in this plot, does require detailed modeling.  See Section~\ref{sec:anal}.
}
\label{fig:caust2}
\end{figure}

Figures~\ref{fig:caust1} and \ref{fig:caust2} illustrate the geometries
of all eight solutions, and
Table~\ref{tab:all} gives the parameters for these solutions.  The notation ($+,-$) is used to indicate solutions where lens-source relative trajectory approaches the caustic from the positive or negative $\theta$ direction in the lens-plane geometry.  Two indices are given for each solution: one for Earth-based observations and one for Spitzer.  
According to Table~\ref{tab:all}, the best wide and close solutions differ
by $\Delta\chi^2<1$.  Hence, this very common degeneracy is completely
unbroken in the present case.  However, since the microlensing parameters
of these two sets of solutions are very similar, their
physical implications are basically the same.

Within each set of solutions (wide or close) the physical implications
are even closer, to the point of being nearly identical.  Therefore,
we present in Table~\ref{tab:brief} a simplified summary of each set
of solutions, with the parameter values being averages over the four
solutions and the ``errors'' taking account of both the fit errors
at different minima and the differences between minima.

{\section{Einstein Radius Estimate}
\label{sec:thetae}}

We estimate $\theta_\e$ using the standard approach \citep{ob03262},
i.e., estimating the source surface brightness from its dereddened 
color $(V-I)_0$ and
the calibrated color/surface-brightness relation of \citet{kervella04},
and then comparing this to dereddened magnitude $I_{s,0}$.  

We begin (as is usual) by assuming that the source is behind the same
dust column as the bulge clump stars.  The field shows substantial differential
reddening, so we restrict to a $60^{\prime\prime}$ radius about the source,
wherein the clump appears compact.  Using OGLE data, we find that the
source lies $\Delta I=3.15$ mag below the clump, whereas using CTIO data
we find $\Delta I=3.01$.  This difference is typical of the difficulty
in centroiding the clump in the vertical (magnitude) direction.  We adopt
$\Delta I=3.08\pm0.07$ and so $I_{s,0} = I_{{\rm cl},0} + \Delta I = 17.54$
where $I_{{\rm cl},0}=14.41$ is adopted from \citet{nataf13}.

Determining $(V-I)_s$ (and so $(V-I)_{s,0}$) presents greater challenges
given that the source is quite faint at baseline $V\sim 21.9$ and that
the full moon was passing through the bulge when the source was most highly
magnified.  OGLE systematically acquires $V$-band data for all its fields,
but of course avoids observing near
the Moon.  Hence, the OGLE $(V-I)_0$ color estimate 
has relatively large errors: $\Delta(V-I)_s = -0.29\pm 0.06$.  CTIO
deliberately targeted the event near peak (and so Moon passage) in 
order to obtain high signal-to-noise measurements.  Some of these observations
were
corrupted by scattered moonlight, but most appear fine.  These data
yield $\Delta(V-I)_s = -0.35\pm 0.03$.  We combine these OGLE and CTIO
measurements by standard
error weighting and obtain $(V-I)_{s,0} = (V-I)_{{\rm cl},0} + \Delta (V-I)=0.72$
where we have adopted $(V-I)_{{\rm cl},0} =1.06$ from \citet{bensby13}.
If we assume that the source is exactly at the
distance to the clump and adopt $M_{I,\rm cl}=-0.12$, 
then this implies $[M_I,(V-I)_0]_s = (2.96,0.72)$.  This is a plausible
pair of values for a star just entering the sub-giant branch from the
turnoff.  That is, the assumption that the source lies behind all the
dust is self-consistent, since it implies that the source inhabits a
reasonably well-populated part of the color-magnitude diagram.

However, another interpretation is that the source lies on the upper
main sequence (since its color is similar to the Sun).  Then,
$M_I\sim 4.15$, so that the star lies 1.2 mag in front of the Bulge in
distance modulus, or at $D_S\sim 4.5\,\kpc$ from us.  In this case,
it would still lie about 125 pc below the Galactic plane and so
behind most but possibly not all the dust toward the Galactic bulge.

We first estimate the source radius under the assumption that
it is a bluish bulge subgiant and then address how this estimate
may be affected if the source is in fact a disk main sequence star.
After using \citet{bb88} to convert $(V-I)$
to $(I-K)$, the above procedure leads to estimates,
\begin{equation}
\theta_* = 1.07\pm 0.10\,\muas;
\qquad
\theta_\e = {\theta_*\over\rho}=0.76\,\mas;
\qquad
\mu_\geo = {\theta_\e\over t_\e} = 4.8\,\masyr .
\label{eqn:thetaeval}
\end{equation}

If the source were a disk main sequence star, but nevertheless lay
behind all the dust, then these estimates would not be affected at all.
Only the dereddened color and magnitude enter the calculation, and these
would not change.  If the source does lie in front of some of the
dust, then it is both intrinsically redder in $(V-I)$ and fainter in $I$
than the above calculations would imply.  Of course fainter stars have
smaller radii at fixed surface brightness, but redder stars have
lower surface brightness (so larger radii) at fixed magnitude.  Hence,
these two effects tend to cancel.  Since the total amount of dust
behind the source is unlikely to be large and the two effects tend
to cancel, and since at this point we cannot reliably estimate how
much dust does lie behind the source, we will proceed on
the assumption that the source is behind all of the dust.
However,
in Section~\ref{sec:discuss}, we discuss how this issue could be
partly or fully resolved in the future.

{\section{Physical Parameters}
\label{sec:physical}}

The extraction of physical parameters is made simpler by the
``resolution'' (or rather irrelevance) of the ambiguity
of the amplitude of $\Delta\beta$ (Section~\ref{sec:anal}),
but more complicated by the ambiguity in the source distance
(Section~\ref{sec:thetae}).  We begin therefore with the
lens mass estimate, which does not depend on the source distance
\begin{equation}
M = {\theta_\e\over\kappa\pi_\e}= 0.38\pm 0.04\,M_\odot;
\qquad
m_{\rm planet} = qM = 21\pm 2 M_\oplus
\label{eqn:mass}
\end{equation}
Similarly, for the projected velocity in the geocentric and heliocentric frames
\begin{equation}
\tilde\bv_\geo(N,E) = {\au\over t_\e}\,{\bpi_\e\over\pi_\e^2} = (0,-124)\kms
\qquad
\tilde\bv_\hel = \tilde\bv_\geo + \bv_{\oplus,\perp} = (0,-95)\kms
\label{eqn:tildev}
\end{equation}
where $\bv_{\oplus,\perp}(N,E)=(-0.8,28.8)\kms$ is Earth's velocity projected
on the plane of the sky at the peak of the event and where we have ignored
the slight differences among the four wide solutions.

Regardless of the source location, the relative parallax is well-determined
\begin{equation}
\pi_\rel =\theta_\e\pi_\e = 0.19\,\mas,
\qquad
\label{eqn:pirel}
\end{equation}
but depending on the distance to the source,
this leads to two different distance estimates for the lens
$D_L = 3.3\,\kpc$ (bulge source) or $D_L = 2.5\,\kpc$ (disk source).
However, within the context of our program of determining the
Galactic distribution of planets, these two outcomes are basically
similar: foreground disk lens.  In Section~\ref{sec:discuss}, 
we discuss how this ambiguity may be resolved with future observations.

Finally, the ambiguity in $D_L$ gives rise to an ambiguity of the same
size in the star-planet projected separation 
$r_\perp=s\theta_\e D_L = 2.7\,\au$ (bulge source) or $2.1\,\au$ (disk source).
Adopting a snowline $r_{\rm snow} = 2.7\,\au\,(M/M_\odot)$, these separations
are at $r_\perp/r_{\rm snow} = 2.2$ and 1.7 snow-line distances, respectively.
Hence, in either case, this planet is a ``cold Neptune''.  Note that
in making these estimates, we have adopted $s=1$, i.e., midway between
(and 10\% different from) the close and wide solutions.

{\section{Discussion}
\label{sec:discuss}}

This is the second microlensing planet (after OGLE-2014-BLG-0124Lb),
whose mass and distance have been characterized with the aid of
parallax observations using {\it Spitzer}.  However, it is the first
to be so characterized as part of a program specifically designed to
measure the Galactic distribution of planets by means of well-defined
selection criteria \citep{yee15}.  While naively such well-defined
criteria might seem to lead to clear cut observing decisions, the
example of OGLE-2015-BLG-0966 demonstrates that they in fact create a
set of complex interlocking constraints, both hard and soft, on
decision processes of multiple semi-autonomous decision makers working
toward a common goal.  Specifically, there was the struggle to
determine whether surveys would cover this event, 
including both static but
non-obvious (KMT) information and dynamic OGLE information, impact on
balancing with coverage of other events (also impacted by survey
coverage), and wild oscillations in {\it Spitzer} cadence due to objective
criteria.  See Section~\ref{sec:obs}.

In the present case, these follow-up observations that were subject
to these interlocking constraints led to detection
and characterization of a planet while preserving the integrity of the
selection process.  In subsequent papers we will show that similar
sensitivity to planets (while maintaining sample integrity) was achieved
under similar conditions for several other high-magnification events.

The character of the planet, continues to confirm that
``cool Neptunes are common'' \citep{ob05169}, even though  that
original claim was based on just two detections.  However, the central
focus of the present effort is not the frequency of planets as a function
of mass, but of Galactocentric radius.  Nothing can be said so far about
this in part because there are so far only two planets in the sample,
but mainly because the sensitivity of the surveys to planets as a function
of Galactocentric radius has not yet been determined (although see
\citealt{zhu15} for initial work in that direction).  Nevertheless, it
is intriguing to note that both detected planets are in the near-to-mid
disk.

While the lens is clearly in the disk, it is unknown at this point
whether the source is in the bulge or the disk.  While the source
distance is of overall secondary interest, it does affect the lens
distance $D_L$ and planet-star separation $r_\perp$ at the 25\% level
and therefore should be resolved if possible.  We note that the
heliocentric projected velocity $\tilde\bv_\hel$ does tend to favor
a disk source just because the solutions all have the heliocentric
projected velocities
pointing basically out of the Galactic plane, and indeed somewhat 
retrograde.  By contrast, as expected naively
(and confirmed by \citealt{21event}) events with bulge sources and
disk lenses tend to have $\tilde\bv_\hel$ aligned with Galactic rotation.
This is because the Sun and the lens both basically partake of this
rotation while bulge source proper motions tend to be both
isotropically distributed and of low amplitude.

The situation would be greatly clarified by measuring the proper motion
of the source, as \citet{ob140939} did for the {\it Spitzer} event
OGLE-2014-BLG-0939.  In contrast to that case, however, in which the source
was bright and easily distinguished from neighbors, which enabled
precise astrometric measurements from over a decade of OGLE observations,
the OGLE-2015-BLG-0966
source star is both too faint and too close $(0.7^{\prime\prime})$
to a neighbor for reliable astrometry.  However, using difference
image astrometry applied to high-magnification images, 
the current position of the lens is measured with a precision of
about 20 mas.  Hence, a single epoch of high-resolution imaging
10 years from now should be able to reliably determine whether the
source proper motion is consistent with disk motion (typically about
$6\,\masyr$).

\acknowledgments

This work is based in part on observations made with the {\it Spitzer} Space Telescope,
which is operated by the Jet Propulsion Laboratory, California Institute of Technology under a contract with
NASA.

The OGLE project has received funding from the National Science Centre,
Poland, grant MAESTRO 2014/14/A/ST9/00121 to AU

Work by JCY, AG, and SC was supported by JPL grant 1500811.  Work by JCY was
performed under contract with the California Institute of Technology
(Caltech)/Jet Propulsion Laboratory (JPL) funded by NASA through the
Sagan Fellowship Program executed by the NASA Exoplanet Science
Institute

Work  by  C.H. was  supported  by  Creative  Research  Initiative Program
(2009-0081561) of National Research Foundation  of  Korea.

GD acknowledges Regione Campania for support from POR-FSE Campania 2014-2020. 

TS acknowledges the financial support from the JSPS, JSPS23103002,JSPS24253004 and JSPS26247023. The MOA project is supported by the grant JSPS25103508 and 23340064.  The US portion of the MOA Collaboration acknowledges financial support from
the NSF (AST-1211875) and NASA (NNX12AF54G).

Work by YS was supported by an
appointment to the NASA Postdoctoral Program at the Jet
Propulsion Laboratory, administered by Oak Ridge Associated
Universities through a contract with NASA.

This publication was made possible by NPRP grant \# X-019-1-006 from the Qatar National Research Fund (a member of Qatar Foundation)

S.D. is supported by “the Strategic Priority Research Program—The
Emergence of Cosmological Structures” of the Chinese Academy of
Sciences (grant No. XDB09000000).

Work by SM has been supported by the Strategic Priority Research Program ``The Emergence of Cosmological Structures" of the Chi-
nese Academy of Sciences Grant No. XDB09000000,
and by the National Natural Science Foundation of
China (NSFC) under grant numbers 11333003 and 11390372.

M.P.G.H. acknowledges support from the Villum Foundation.  Based on data collected by MiNDSTEp with the Danish 1.54 m telescope at the ESO La Silla observatory. JSurdej and OW acknowledge support from the Communauté française de Belgique – Actions de recherche concéertes – Académie Wallonie-Europe. SHG and XBW would like to thank the financial support from National Natural Science Foundation of China through grants Nos. 10873031 and 11473066. 

This work makes use of observations from the LCOGT network, which
includes three SUPAscopes owned by the University of St Andrews.  The
RoboNet programme is an LCOGT Key Project using time allocations from
the University of St Andrews, LCOGT and the University of Heidelberg
together with time on the Liverpool Telescope through the Science and
Technology Facilities Council (STFC), UK. This research has made use
of the LCOGT Archive, which is operated by the California Institute of
Technology, under contract with the Las Cumbres Observatory.

\begin{deluxetable}{lrrrrrrrr}

\rotate

\centering

\tablecaption{All solutions.}

\tabletypesize{\scriptsize}

\tablehead{Parameters & $(+,+)$,wide & $(+,-)$,wide & $(-,+)$,wide & $(-,-)$,wide & $(+,+)$,close & $(+,-)$,close & $(-,+)$,close & $(-,-)$,close}

\startdata

$\chi^2$/dof & 15136/15278 & 15138/15278 & 15137/15278 & 15135/15278 & 15136/15278 & 15151/15278 & 15154/15278 & 15135/15278 \\
$t_0$ & 7205.1979 & 7205.1953 & 7205.1955 & 7205.1937 & 7205.1978 & 7205.1945 & 7205.1936 & 7205.1938 \\
 & 0.0012 & 0.0013 & 0.0013 & 0.0012 & 0.0012 & 0.0012 & 0.0012 & 0.0012 \\
$u_0$ & 0.01136 & 0.01136 & -0.01157 & -0.01152 & 0.01136 & 0.01136 & -0.01162 & -0.01149 \\
 & 0.00010 & 0.00010 & 0.00010 & 0.00010 & 0.00010 & 0.00010 & 0.00010 & 0.00009 \\
$t_{\rm E}$ (days) & 57.7 & 57.8 & 57.7 & 57.7 & 57.7 & 57.7 & 57.5 & 57.8 \\
 & 0.4 & 0.4 & 0.4 & 0.4 & 0.4 & 0.5 & 0.4 & 0.4 \\
$\rho$ ($10^{-4}$) & 14.08 & 14.00 & 14.06 & 14.06 & 14.07 & 14.02 & 14.12 & 14.04 \\
 & 0.14 & 0.15 & 0.14 & 0.13 & 0.15 & 0.15 & 0.15 & 0.14 \\
$\pi_{\rm E,N}$ & 0.0234 & -0.0561 & 0.0397 & -0.0412 & 0.0234 & -0.0538 & 0.0317 & -0.0412 \\
 & 0.0012 & 0.0015 & 0.0012 & 0.0013 & 0.0012 & 0.0059 & 0.0094 & 0.0012 \\
$\pi_{\rm E,E}$ & -0.238 & -0.237 & -0.232 & -0.237 & -0.239 & -0.264 & -0.272 & -0.237 \\
 & 0.006 & 0.012 & 0.003 & 0.006 & 0.006 & 0.007 & 0.007 & 0.006 \\
$\alpha$ (deg) & 50.7 & 50.6 & -50.4 & -50.4 & 50.6 & 50.6 & -50.3 & -50.4 \\
 & 0.1 & 0.1 & 0.1 & 0.1 & 0.1 & 0.1 & 0.1 & 0.1 \\
$s$ & 1.1141 & 1.1146 & 1.1148 & 1.1147 & 0.9093 & 0.9091 & 0.9088 & 0.9091 \\
 & 0.0033 & 0.0032 & 0.0033 & 0.0033 & 0.0026 & 0.0025 & 0.0027 & 0.0026 \\
$q$ ($10^{-4}$) & 1.67 & 1.67 & 1.68 & 1.68 & 1.70 & 1.70 & 1.71 & 1.70 \\
 & 0.05 & 0.05 & 0.05 & 0.05 & 0.05 & 0.05 & 0.06 & 0.05 \\
Blend & -0.027 & -0.024 & -0.027 & -0.026 & -0.027 & -0.025 & -0.031 & -0.024 \\
 & 0.008 & 0.009 & 0.008 & 0.008 & 0.008 & 0.008 & 0.008 & 0.008 \\

\enddata
\tablecomments{$^a$HJD-2450000.}
\label{tab:all}
\end{deluxetable}

\begin{table}

\centering

\caption{Abbreviated table of parameter values averaged over the four solutions.}

\begin{tabular}{lrr}

\tableline\tableline

Parameters & Wide solution & Close solution \\

\tableline

Best $\chi^2$ & 15135 & 15135 \\

$t_0^{a}$ & $7205.196\pm0.002$ & $7205.195\pm0.002$ \\

$|u_0|$ & $0.0115\pm0.0001$ & $0.0115\pm0.0001$ \\

$t_{\rm E}$ (days) & $57.7\pm0.4$ & $57.7\pm0.4$ \\

$\rho$ ($10^{-4}$) & $14.04\pm0.15$ & $14.07\pm0.15$ \\

$|\pi_{\rm E,N}|$ & $0.04\pm0.02$ & $0.03\pm0.02$ \\

$\pi_{\rm E,E}$ & $-0.24\pm0.01$ & $-0.25\pm0.02$ \\

$|\alpha|$ (deg) & $50.5\pm0.2$ & $50.5\pm0.2$ \\

$s$ & $1.115\pm0.004$ & $0.909\pm0.003$ \\

$q$ ($10^{-4}$) & $1.68\pm0.05$ & $1.70\pm0.06$\\

\tableline\tableline
$^{a}$HJD-2450000
\end{tabular}
\label{tab:brief}
\end{table}

\end{document}